\documentclass[a4paper,12pt]{article}

\usepackage{hyperref}

\usepackage{amsmath}
\usepackage{graphicx}

\usepackage[margin=1.0in]{geometry}
\usepackage{authblk}

\title{Formation of photon spheres in boson stars\\
   with a nonminimally coupled field}

\author{D.~Horvat}
\author{S.~Iliji\'c\footnote{Email: {\tt sasa.ilijic@fer.hr}}}
\author{A.~Kirin}
\author{Z.~Naran\v ci\'c}
\affil{University of Zagreb, Faculty of Electrical Engineering and Computing,\\
   Department of Applied Physics, Unska 3, HR-10\,000 Zagreb, Croatia}

\date{June 18, 2013}

\begin{document}

\maketitle 


\begin{abstract}
A static, spherically symmetric, asymptotically flat spacetime
may allow for circular, closed null-geodesics
which are said to belong to a photon sphere.
In the context of gravitational lensing
in the strong deflection regime,
the presence of a photon sphere
leads to an unbounded angle of deflection of light
(multiple turns) and formation of relativistic images.
In this paper, we show that photon spheres may form
in some configurations of boson stars
constructed with a free massive complex scalar field
nonminimally coupled to gravity.
Assuming that the boson star is transparent to light,
photon spheres would give raise not only to phenomena
in the realm of strong gravitational lensing,
but also to considerably increased photon flux
in the central region of the star,
relative to the flux in its surroundings.
\end{abstract}


\section{Introduction \label{sec:intro}}


Deflection of light by a massive body,
as predicted by General Relativity, led to one of the first
experimental verifications of the theory \cite{WeinbergBook}.
Deflection of light is also the mechanism
behind the rich phenomenology of gravitational lensing
that has seen a wide range of
applications in astrophysics \cite{LensingLNP}.
In observationally relevant circumstances,
gravitational lensing occurs
in the so-called weak deflection regime;
the light passes by a massive body at distances
that are much greater than the Schwarzschild radius corresponding to the mass,
and the overall angle of deflection is small.
However, despite known difficulties on the observational side,
there is also considerable interest in gravitational lensing
in the strong deflection regime,
see e.g.\ \cite{eiroaSDGL} and references therein.
This primarily includes the case of lensing by the Schwarzschild black hole
\cite{ohanianBHGL,virbhadraellis,virbhadra2009},
by the regular black hole \cite{eiroaBardeen}
or by other types of black holes, see \cite{bozzaReview} for a review.
Light passing sufficiently close to the Schwarzschild black hole may,
due to the presence of the photon sphere in the spacetime,
suffer an arbitrarily large deflection
as it makes one or more complete turns around it
before escaping to spatial infinity.
In the context of gravitational lensing,
this leads to the formation of the so-called relativistic images.
A mathematically rigorous treatment of the concept of the photon sphere
can be found in \cite{ClaudelVirbhadraEllisJMP}.


In this paper, we look for a simple matter model
that can be used to construct static,
spherically symmetric self-gravitating bodies
that curve the spacetime in such a way that photon spheres are formed.
In this regard, one is faced with two possible routes.
One is to construct a compact body,
i.e.\ a body that is finite in the radial extent,
with the surface lying below the photon sphere of the Schwarzschild spacetime.
If such a body is assumed to be non-transparent to photons,
then its properties as a gravitational lens
are equivalent to those of a Schwarzschild black hole.
The other more interesting route
is to allow the energy-momentum content to take up all space
and assume that the matter it describes is transparent to light.
The simplest matter model for which it is natural to assume
that it does not interact electromagnetically is the scalar field.
For example, lensing by the configurations
of a massless scalar field with a naked spacetime singularity
was studied in \cite{virbhadraScalar}.
As we intend to show,
photon spheres can form within the regular objects
formed of scalar fields known as boson stars.


Boson stars were originally conceived as Klein--Gordon geons
\cite{KaupKleinGordonGeon,RuffiniBonazzola},
the self-gravitating configurations
of the free massive scalar field minimally coupled to gravity,
while among the extensions of the original model one finds
a quartic self-interaction \cite{CoShaWa86},
nonminimal coupling to gravity $\cite{vdbijgleiser,horvatmarunovic}$,
addition of the Brans--Dicke field \cite{BransDickeBS}, etc.
Due to their simplicity,
boson stars proved to be an attractive theoretical model
for studying the properties of self-gravitating objects in General Relativity.
They were also considered as candidates for astrophysical objects
such as supermassive Sgr~$\mathrm{A}^*$
at the centre of our galaxy \cite{torres2000}.
Several reports on the subject appeared over the past decades
\cite{jetzerReport,SchuMi03,LiebPalenLRev}.
Gravitational lensing by boson stars
was considered in \cite{DabroSchunck00,binnunay};
relatively large angles of deflection of light
forming primary and secondary images were found,
but the possibility of formation of photon spheres
and relativistic images was not reported so far.


This paper is organized as follows.
In Section \ref{sec:notation}, we give a brief overview
of the properties of null-geodesics in a static,
spherically symmetric, asymptotically flat spacetime with photon spheres
and introduce the notation that will be used throughout the rest of the paper.
We also discuss a simple analytical model of a compact body
with the surface within the Schwarzschild photon sphere.
In Section \ref{sec:bstar}, we construct boson stars from
a free massive complex scalar field nonminimally coupled to gravity
and show that photon spheres form in some configurations.
We sum up in Section \ref{sec:concl}.


\section{Notation and a simple model \label{sec:notation}}


We will be considering the behaviour of null-geodesics in static,
spherically symmetric, asymptotically flat spacetimes,
free of singularities or black holes.
The line element can be written
using the coordinates $x^a = (t,r,\vartheta,\varphi)$
and geometrized units ($c=1=G$) as
   \begin{equation} \label{eq:ds2}
   \mathrm{d} s^2 = g_{ab} \, \mathrm{d}x^a \, \mathrm{d}x^b
                  = - \mathrm{e}^{2\Phi(r)} \mathrm{d}t^2
                    + \mathrm{e}^{2\Lambda(r)} \mathrm{d}r^2
                    + r^2 \mathrm{d}\Omega^2 ,
   \end{equation}
where the radial coordinate $r$ is the area radius
and $ \mathrm{d}\Omega^2 = \mathrm{d} \vartheta^2
                         + \sin^2\vartheta \, \mathrm{d} \varphi^2 $
is the line element on the unit sphere.
The Schwarzschild spacetime
contains a black hole of mass $M>0$ with the event horizon at $r=2M$.
Outside the black hole,
the metric components of the Schwarzschild spacetime are given by
   \begin{equation} \label{eq:sch}
   \mathrm{e}^{2\Phi(r)} = \mathrm{e}^{-2\Lambda(r)} = 1 - {2M}/{r},
   \qquad r>2M,
   \end{equation}
$M$ being the ADM mass of the spacetime.
We will encounter the situation
where the the energy-momentum tensor vanishes
at radii greater than some constant $R>2M$
and where the metric components are given by (\ref{eq:sch}) exactly,
as well as situations where nonzero energy-momentum takes up all space
and where (\ref{eq:sch}) only represents
the asymptotic form of the metric as $r\to\infty$.


The Lagrangian 
for the geodesics constrained to the $\vartheta=\pi/2$ plane
can be written as
   \begin{equation} \label{eq:geolag}
   L = \sqrt{\pm \left(- \mathrm{e}^{2\Phi(r)} \dot t^2
       + \mathrm{e}^{2\Lambda(r)} \dot r^2
       + r^2 \dot\varphi^2 \right)},
   \end{equation}
where the overdot denotes differentiation
with respect to the parameter of the geodesic
and the choice of the sign
depends on whether the geodesic is spacelike ($+$) or timelike ($-$).
As $L$ does not depend on $t$ and $\varphi$, there are two constants of motion,
${\partial L}/{\partial \dot t}$ and ${\partial L}/{\partial \dot \varphi}$.
The ratio of the two constants,
   \begin{equation} \label{eq:bdef}
   b = \frac{\partial L/\partial \dot \varphi}{\partial L/\partial \dot t}
     = - r^2 \mathrm{e}^{-2\Phi(r)} \frac{\mathrm{d}\varphi}{\mathrm{d}t}
     = \mathrm{const.},
   \end{equation}
is proportional to the angular momentum
of the test particle following a timelike geodesic
and is known as the impact parameter.
For a geodesic that can be extended to spatial infinity,
i.e.\ to the region where the spacetime is essentially flat,
the absolute value of the impact parameter
can be interpreted as the distance
between the considered geodesic and the purely radial geodesic
that is (at spatial infinity) parallel to it.


By continuity, the impact parameter $b$ defined in (\ref{eq:bdef})
is a constant of motion also for the null-geodesics,
where in addition we have $L=0$.
Eliminating $\mathrm{d}\varphi/\mathrm{d}t$
one obtains an equation for $\mathrm{d}r/\mathrm{d}t$
that can be written in the form of the nonrelativistic energy equation,
   \begin{equation} \label{eq:drdt}
   \frac12 \left(\frac{\mathrm{d}r}{\mathrm{d}t}\right)^2 = E - V(b,r),
   \end{equation}
where $E=0$ and the `effective potential' $V(b,r)$ is given by
   \begin{equation} \label{eq:vdef}
   V(b,r) = \frac{\mathrm{e}^{2\Phi(r)-2\Lambda(r)}}2
            \left( \frac{b^2 \mathrm{e}^{2\Phi(r)}}{r^2} - 1 \right).
   \end{equation}
In analogy with the familiar scenario from classical mechanics,
we see that null-geodesics exist only where $V(b,r) \le 0$,
as the rhs of (\ref{eq:drdt}) must be non-negative
for $\mathrm{d}r/\mathrm{d}t$ to be real.
If $V(b,r_0)=0$, then at $r=r_0$,
there is a turning point of $r(t)$
   if $\partial V(b,r) / \partial r |_{r=r_0}\ne0$,
or an equilibrium point 
   if $\partial V(b,r) / \partial r |_{r=r_0}=0$.
A turning point is an inner turning point (a minimum of $r(t)$)
if $\partial V(b,r) / \partial r |_{r=r_0}< 0$;
it is an outer turning point (a maximum of $r(t)$)
if $\partial V(b,r) / \partial r |_{r=r_0}> 0$.
An equilibrium point of $r(t)$ implies the existence of closed
(circular) null-geodesics along equators of the hypersurface $r=r_0$
which is known as the photon sphere.
A photon sphere is said to be stable
if $\partial^2 V(b,r) / \partial r^2 > 0$,
or unstable if $\partial^2 V(b,r) / \partial r^2 < 0$.
In the neighborhood of a stable photon sphere,
there are bound null-geodesics with oscillatory $r(t)$
such that the outer turning point is above
and the inner turning point is below the photon sphere.
Apart from circular null-geodesics,
an unstable photon sphere implies the existence of null-geodesics
that asymptotically approach the photon sphere,
either from the inner or from the outer side,
making an infinite number of turns around it.
It can be shown that the outermost photon sphere
in an asymptotically flat, static, spherically symmetric spacetime
is an unstable photon sphere.


To access the properties of null-geodesics
in a given static spherically symmetric spacetime 
in a systematic way
it is convenient to consider the relation among the impact parameter $b$
and the radial coordinates of turning points or photon spheres $r_0$.
This relation follows from the condition
$ \mathrm{d} r / \mathrm{d} t |_{r=r_0} = 0$
or $V(b,r_0)=0$, and reads
   \begin{equation} \label{eq:br0}
   b(r_0) = \pm r_0 \mathrm{e}^{-\Phi(r_0)}.
   \end{equation}
The double sign in (\ref{eq:br0})
corresponds to the sign of the angular momentum
or the time direction one chooses along the given null-geodesic.
For simplicity of notation, with no loss in generality,
we will assume a positive $b$ in what follows.
It can be easily shown that solutions to (\ref{eq:br0})
with ${\mathrm{d}b}/{\mathrm{d}r_0}>0$ represent inner turning points,
while those with ${\mathrm{d}b}/{\mathrm{d}r_0}<0$
represent outer turning points of null-geodesics;
the solutions with ${\mathrm{d}b}/{\mathrm{d}r_0} = 0$
and ${\mathrm{d}^2b}/{\mathrm{d}r_0^2}<0 $ represent stable,
while those $ {\mathrm{d}^2b}/{\mathrm{d}r_0^2} > 0 $
represent unstable photon spheres in the spacetime.

\begin{figure}
\begin{center}
\includegraphics{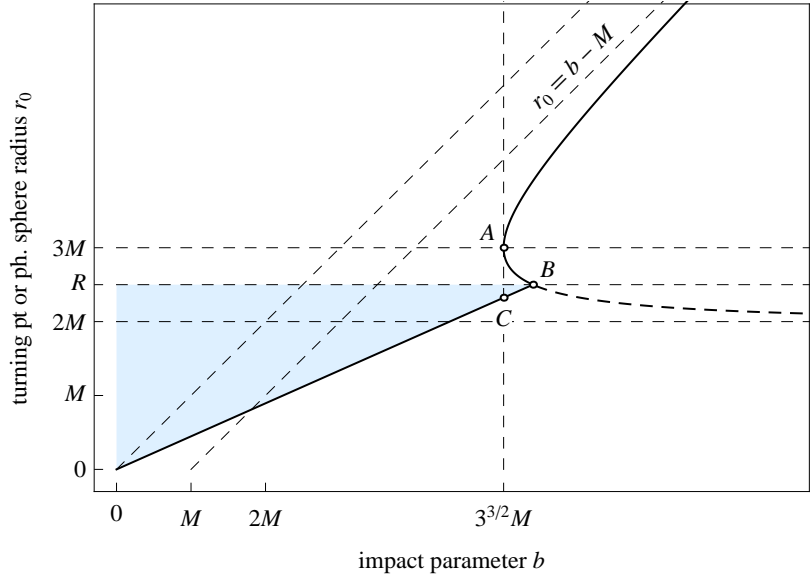}
\end{center}
\caption{\label{fig:one}
Turning point or photon sphere radius $r_0$
versus impact parameter $b$ (thick solid line)
for null-geodesics in the spacetime containing
the thin spherical shell of mass $M$ and radius $R=\frac52 M$.
(Null-geodesics exist only to the left of the thick solid line;
the shaded region corresponds to the shell interior.)
The thick dashed line is the continuation of the $r_0$ versus $b$ curve
for the case of the black hole of mass $M$.
Thin dashed lines are provided as eye-guides;
the line $r_0=b-M$ is the asymptote of $b(r_0)$ for $b\to\infty$.}
\end{figure}

Let us first consider the region of the Schwarzschild spacetime
outside the black hole
where the metric components are given by (\ref{eq:sch}).
There is no solution to (\ref{eq:br0})
with $r_0>0$ for the impact parameter $b<3^{3/2}M$.
This implies that null-geodesics with $b<3^{3/2}M$
extend to spatial infinity at one end,
while their other end reaches the black hole horizon.
At $b=3^{3/2}M$, there is the unstable photon sphere
with an area radius $r_0=3M$ (see point $A$ in figure~\ref{fig:one}).
For $b>3^{3/2}M$, one finds two turning points.
The inner turning point with $r_0>3M$
and with the asymptote $r_0=b-M$ as $b\to\infty$
represents the closest approach
of the null-geodesics extending to spatial infinity.
The outer turning point with $r_0<3M$
and with the asymptote $r_0=2M$ as $b\to\infty$
represents the outermost point reached
by the geodesics with both ends at the black hole horizon.


We now proceed to replace
the central segment of the Schwarzschild spacetime
containing the black hole of mass $M$
with an exact solution of the Einstein equations
representing a static spherical body (compact object)
of mass $M$ and surface radius $R>2M$.
We choose to do so in such a way that the joining hypersurface $r=R$
lies inside the Schwarzschild photon sphere, i.e.\ we require $2M<R<3M$.
This implies that the ratio $2M/R$,
known as the surface compactness of the body,
belongs to the highly relativistic regime,
   \begin{equation} \label{eq:mu23}
   \frac23 < \frac{2M}{R} < 1.
   \end{equation}
Here, one recalls  Buchdahl's upper bound on the surface compactness
for bodies involving isotropic pressures in their interior \cite{Buch59},
$2M/R\le8/9$,
and also that many stellar models develop disturbing features
such as violation of the energy conditions
or loss of dynamical stability at high values of surface compactness.
However, simple models of massive bodies
with sufficiently high surface compactness
to satisfy condition (\ref{eq:mu23}) can be found,
and we choose here to work with a model consisting
of a massive spherical shell of infinitesimal thickness
supported against gravity by its own surface pressure.
It was shown in \cite{FrauHoenKon90}
that for the case of the flat metric in the shell interior,
this model allows the surface compactness as high as $24/25$
without the violation of the dominant energy condition.
The metric components inside the shell are given by
   \begin{equation} \label{eq:int}
   \mathrm{e}^{2\Phi(r)} = 1 - 2M/R, \qquad
   \mathrm{e}^{2\Lambda(r)} = 1 , \qquad r<R.
   \end{equation}
It is important to note that joining metrics
(\ref{eq:sch}) and (\ref{eq:int}) at $r=R$ is not smooth.
While $g_{tt} = - \mathrm{e}^{2\Phi(r)}$ is continuous,
$g_{rr} = \mathrm{e}^{2\Lambda(r)}$ has a discontinuity at $r=R$.
This feature is closely related
to the $\delta$-shaped distribution of matter of the massive shell
along the radial coordinate.
The full technical details of the construction
of the energy-momentum tensor on the hypersurface $r=R$
which involves the application
of Israel's thin shell formalism \cite{Israel66}
are not needed for this discussion;
we refer the interested reader to \cite{FrauHoenKon90}
or to \cite{VissGS1,gravaec} where a similar model with
a segment of the de Sitter spacetime in the interior,
known as the gravastar, is worked out in detail.

\begin{figure}
\begin{center}
\includegraphics{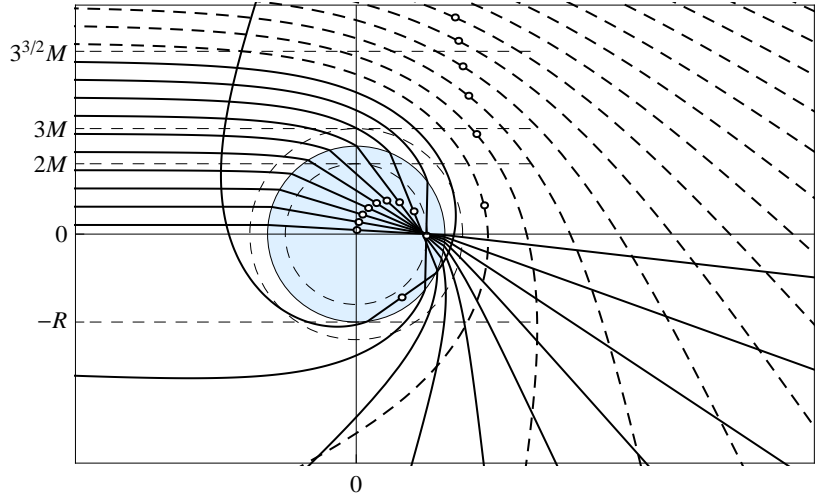}
\end{center}
\caption{\label{fig:two}
A bundle of asymptotically parallel null-geodesics
passing by (impact parameter $b>3^{3/2}M$, dashed lines)
or through ($b<3^{3/2}M$, solid lines)
the spherical shell of mass $M$ and radius $R=\frac52 M$
(thin solid circle with shaded interior).
The turning point of each null-geodesic
is indicated with the small circle.}
\end{figure}

The relation between the impact parameter $b$
and the radial coordinate of the turning point or a photon sphere $r_0$
for the spacetime containing the massive shell obeying (\ref{eq:mu23})
is shown in figure~{\ref{fig:one}}.
The curved line is the plot of (\ref{eq:br0})
for the Schwarzschild metric (\ref{eq:sch})
which is relevant for $r_0>R$
(the part of this curve below the point $B$ is shown dashed).
The straight line with one end at $b=r_0=0$ and the other at $B$
is the plot of (\ref{eq:br0}) for the metric inside the shell (\ref{eq:int})
which is relevant for $r_0<R$.
The point $A$ with the coordinates $r_0=3M$ and $b=3^{3/2}M$
is the Schwarzschild photon sphere,
while the part of the curve above $A$
represents the inner turning points (radii of closest approach)
for null-geodesics with $b>3^{3/2}M$ reaching spatial infinity.
These null-geodesics are entirely contained
in the Schwarzschild segment of the spacetime
and are in no way affected by the massive shell replacing the black hole.
Part of the curve between $A$ and $B$
   represents the outer turning points,
and the part of the straight line between $C$ and $B$
   represents the inner turning points,
for bound null-geodesics with the impact parameter $b$ in the range
$3^{3/2}M < b < b_B = {R}/{\sqrt{1-2M/R}}$.
Part of the straight line below $C$ represents
the inner turning points (radii of closest approach)
for null-geodesics reaching spatial infinity with $b<3^{3/2}M$.
The turning points of these null-geodesics
lie within the sphere of area radius
   \begin{equation}
   r_{C} = 3^{3/2} M \sqrt{1-2M/R},
   \end{equation}
which is smaller than the radius of the shell.
It is easy to see that $r_{\mathrm{C}}\to0$ for $2M/R\to1$.
This also means that there is no null-geodesic extending to spatial infinity
with the radius of closest approach $r_0$ in the range $r_{C} \le r_0 \le 3M$.
The behaviour of a bundle of null-geodesics
that are parallel at spatial infinity is shown in figure~\ref{fig:two}
as it reaches the massive shell of radius $R=5M/2$.
The null-geodesics with the impact parameter $b<3^{3/2} M$ (solid lines)
show a strong degree of focusing in the interior of the massive shell.
This interesting feature could not be confirmed analytically
since the null-geodesics in the region $r>R$ could only be constructed
by numeric integration of (\ref{eq:bdef}) and (\ref{eq:drdt}).


In summary of the model we have considered,
we refer to the points $A$, $B$ and $C$ in figure~\ref{fig:one}.
Point $A$ is the outer (unstable) photon sphere,
while $B$ is the inner (stable) photon sphere.
Point $C$ is defined so that $b_C = b_A$.
The null-geodesics extending to spatial infinity
have turning points with $r_0 < r_{C}$ for the impact parameter $b<b_A$,
                 or with $r_0 > r_{A}$ for $b>b_A$.
In addition, there is a class of bound null-geodesics
with impact parameter $b$ in the range $b_A<b<b_B$
and the turning points $r_0$ in the range $r_{C} < r_0 < r_{A}$.
All these qualitative features,
i.e.\ the points $A$, $B$ and $C$ arranged as in figure~\ref{fig:one},
are to be found in a different context in the following section.


\section{Photon spheres in boson stars \label{sec:bstar}}


The action involving the massive complex scalar field $\phi$
nonminimally coupled to gravity is
   \begin{equation} \label{eq:action}
   S = \int \mathrm{d}^4 x \sqrt{-g} \left(
   \left( \frac{1}{16\pi} + \xi \phi^*\phi \right) R
   - (\nabla^a \phi^*)(\nabla_a \phi)
   - \mu^2 \phi^*\phi 
   \right),
   \end{equation}
where $\xi$ is the nonminimal coupling constant,
$R$ is the curvature scalar and $\mu$ is the field mass.
Since the action is invariant under the global phase transformation
$\phi \to \mathrm{e}^{\mathrm{i}\epsilon} \phi$,
there is the conserved current,
$j_a = \mathrm{i} ( (\nabla_a\phi^*) \phi - (\nabla_a\phi) \phi^* )$,
and the corresponding generator, the particle number $N$.
The variation of (\ref{eq:action}) with respect to the metric
gives the Einstein equation, $G_{ab} = 8\pi T_{ab}$,
where $G_{ab} = R_{ab} - \frac12g_{ab}R$ is the Einstein tensor
and the energy--momentum tensor is given by
   \begin{alignat}{1} \label{eq:tabgen}
   T_{ab} & = \nabla_a \phi^* \nabla_b \phi + \nabla_a \phi \nabla_b \phi^*
              - g_{ab} \left(\nabla^c \phi^* \nabla_c \phi
              + \mu^2 \phi^* \phi 
                                               \right) \notag \\
          & \qquad - 2 \xi \, \phi^* \phi G_{ab}
                   - 2 \xi \, g_{ab} \nabla^c\nabla_c \phi^*\phi
                   + 2 \xi \, \nabla_a \nabla_b \phi^*\phi .
   \end{alignat}
Note that the energy--momentum tensor can be written in a different way
if the term $2 \xi \phi^* \phi G_{ab}$
is transferred to the lhs of the Einstein equation,
which is then divided by $1 + 16\pi \xi \phi^* \phi$.
There are also different approaches
to the interpretation of the energy-momentum tensor
in the case of the nonminimally coupled scalar field;
for a discussion see \cite{belluccifaraoni}.


In spherical symmetry, with the metric written as in (\ref{eq:ds2}),
the nontrivial components of the Einstein tensor are
\begin{alignat}{1}
   G^t{}_t & = r^{-2} \big( \mathrm{e}^{-2\Lambda}(1-2r\Lambda')-1\big) ,
      \label{eq:eintentt} \\
   G^r{}_r & = r^{-2} \big( \mathrm{e}^{-2\Lambda}(1+2r\Phi')-1\big) ,
      \label{eq:eintenrr} \\
   G^\vartheta{}_\vartheta = G^\varphi{}_\varphi & =
      r^{-2} \mathrm{e}^{-2\Lambda} \big(
         (r\Phi' - r\Lambda')(1 + r\Phi') + r^2 \Phi'' \big)
             \label{eq:eintenang}
\end{alignat}
(explicit notation of the $r$-dependence of functions
is omitted and the prime denotes the $r$-derivatives).
With the stationary Ansatz for the complex field,
$ \phi = \phi(r) \mathrm{e}^{-\mathrm{i}\omega t} $,
where $\phi$ from here on denotes the real function,
the nontrivial components of the energy-momentum tensor
can be identified with the energy density $\rho$,
radial pressure $p$ and the transverse pressure $q$.
They are given by
\begin{alignat}{1}
   \rho = - T^t{}_t
          & = (\mu^2 + \mathrm{e}^{-2\Phi}\omega^2) \phi^2 
              + \mathrm{e}^{-2\Lambda} \phi'^2
              + 2 \xi \phi^2 G^t{}_t \notag \\
          & \qquad + 4\xi \mathrm{e}^{-2\Lambda} \big(
                 \phi'^2 + (2/r-\Lambda')\phi'\phi + \phi''\phi
              \big), \label{eq:rho} \\[3pt]
   p = T^r{}_r
          & = (-\mu^2 + \mathrm{e}^{-2\Phi}\omega^2) \phi^2 
            + \mathrm{e}^{-2\Lambda} \phi'^2 - 2 \xi \phi^2 G^r{}_r \notag \\
          & \qquad - 4\xi \mathrm{e}^{-2\Lambda}
            (2/r+\Phi')\phi'\phi, \label{eq:ppp} \\[3pt]
   q = T^\vartheta{}_\vartheta = T^\varphi{}_\varphi
            & = (-\mu^2 + \mathrm{e}^{-2\Phi}\omega^2) \phi^2 
            - \mathrm{e}^{-2\Lambda} \phi'^2
            - 2 \xi \phi^2 G^\vartheta{}_\vartheta \notag \\
          & \qquad - 4\xi \mathrm{e}^{-2\Lambda} \big(
            \phi'^2 + (1/r-\Lambda'+\Phi')\phi'\phi + \phi''\phi \big).
               \label{eq:qqq}
\end{alignat}
Plugging (\ref{eq:eintentt})--(\ref{eq:eintenang})
and (\ref{eq:rho})--(\ref{eq:qqq}) into the Einstein equation
gives the system of three coupled ordinary differential equations
with three unknown functions: $\Phi$, $\Lambda$ and $\phi$.
The differential order of the system equals 5
since the highest derivatives that occur
are $\Phi''$, $\Lambda'$ and $\phi''$.
The conservation condition $\nabla_a T^a{}_b = 0$, which gives
   \begin{alignat}{1} \label{eq:conslaw}
   & r^2 \mathrm{e}^{2\Lambda} (-\mu^2+\mathrm{e}^{-2\Phi}\omega^2) \phi^2
     + (2-r\Lambda'+r\Phi')r\phi'\phi + r^2\phi''\phi \notag \\
   & \qquad = 2\xi \big(1-\mathrm{e}^{2\Lambda} + 2r\Phi' + r^2\Phi'^2
              - r\Lambda'(2+r\Phi') + r^2\Phi''\big) \phi^2,
   \end{alignat}
can be used to eliminate $\Phi''$ from the system.
This reduces the differential order of the system by 1.
(Note that (\ref{eq:conslaw}) is equivalent
to what one obtains from (\ref{eq:action})
for the equation of motion for the scalar field.)
The solution is determined with the boundary conditions
which reflect the requirements that all functions are regular at $r=0$,
that the metric is asymptotically flat and coincides with (\ref{eq:sch})
and that the field vanishes as $r\to\infty$.
The number of the required boundary conditions
equals the differential order of the system (four)
plus one since the system also involves the unknown frequency $\omega$
(eigenvalue) which has the role of an additional degree of freedom.
The particle number is given with
   \begin{equation}
   N = \int \mathrm{d}^3 x \, \sqrt{-g} \; j^0
     = \int_0^{\infty} 8\pi r^2 \mathrm{e}^{\Lambda-\Phi} \omega \phi^2
       \, \mathrm{d}r
   \end{equation}
and can be obtained once the functions $\Phi$, $\Lambda$ and $\phi$,
and the frequency $\omega$ have been determined.


In order to set up the numerical procedure, several further steps are taken.
It is convenient to replace the metric function $\Lambda(r)$
with the so called `mass function' $m(r)$ defined with
   \begin{equation} \label{eq:massfn}
   g_{rr} = \mathrm{e}^{2\Lambda} = \frac1{1-2m/r}
   \end{equation}
(in the non-relativistic regime,
$m$ is the mass within the sphere of radius $r$).
The asymptotic value of $m$ as $r\to\infty$
is the ADM mass of the spacetime, $M$.
The radial coordinate $\tilde x=\mu r$ is introduced
which is then mapped onto the compact domain
with the transformation $x = \tilde x/(1+\tilde x) \in [0,1]$.
Following the usual practice in the literature,
we introduce the field variable $\sigma = \sqrt{8\pi} \phi$.
Finally, for the boundary conditions
at $x=0$ (corresponding to $r=0$),
we choose $m=0$, $\sigma=\sigma_0$ and $\sigma'=0$,
while at $x=1$ (corresponding to $r\to\infty$)
we choose $\Phi=0$ and $\sigma=0$.
The central value of the field variable, $\sigma_0=\sigma(0)$,
is used to parametrize the family of solutions obtained for some $\xi$;
for each $\xi$ and $\sigma_0$,
the solution consists of the three functions, $\Phi$, $m$ and $\sigma$,
and the eigenvalue $\omega$.
We used the {\sc colsys} boundary value code \cite{COLSYS}
and checked that our procedure fully reproduces
the results presented in table~1 of~\cite{vdbijgleiser}.


We limited our search for the photon spheres
in the spacetimes sourced by the boson stars
to the configurations without nodes in $\phi$.
(The configurations with one or more nodes
are sometimes referred to as the excited boson stars \cite{jetzerExcStab}.)
For a given value of $\xi$,
as the central value of the field variable $\sigma_0$
increases starting from $\sigma_0=0$,
the mass $M$ and the particle number $N$ also increase
from zero up to the (first) critical configuration
where they reach their maxima.
In analogy with the similar behaviour of mass in fluid stars,
the first maximum of $M$ suggests the onset of dynamical instability.
In the case of minimal coupling of the scalar field and gravity ($\xi=0$),
the loss of stability with respect to radial perturbations
at the first critical configuration has been confirmed by perturbative analysis
\cite{tdleeypangstab,gleiserwatkins}
and also by other methods \cite{fkusma,hawchop}.
As $\sigma_0$ increases beyond the first critical configuration,
the mass $M$ and the particle number $N$
go through a sequence of minima and maxima
resembling strongly damped oscillations
about respective asymptotic values.
Subsequent maxima of $M$ and $N$,
that will also be referred to as critical configurations,
are lower than their maxima at the first critical configuration.

\begin{figure}
\begin{center}
\includegraphics{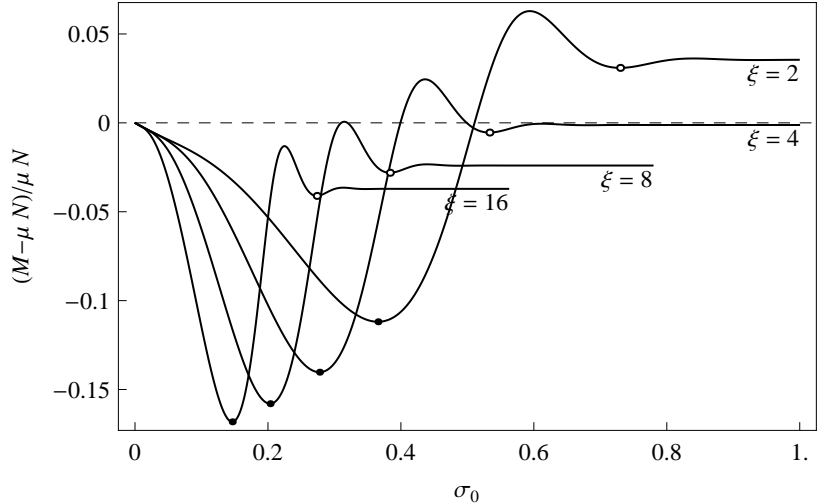}
\end{center}
\caption{\label{fig:three}
Binding energy (\ref{eq:ebind}) of the boson star
per particle in units of the boson mass $\mu$
versus the central value of the field variable $\sigma_0$
for the nonminimal coupling constant $\xi=2,4,8,16$.
First and second critical configurations
are indicated with filled and hollow circles, respectively.
}
\end{figure}

The binding energy of a boson star can be defined as
   \begin{equation} \label{eq:ebind}
   E_{\mathrm{b}} = M - \mu N.
   \end{equation}
For all values of $\xi$ that we considered,
we found that the critical configurations (maxima of $M$ and $N$)
correspond to the minima of $E_{\mathrm{b}}$.
In all the first critical configurations,
the binding energy $E_{\mathrm{b}}$ is negative,
which means that the bosons in such configurations
may not disperse to infinity without the input of additional energy.
For negative $\xi$ and up to $\xi=2$,
as $\sigma_0$ increases beyond the first critical configuration,
$E_{\mathrm{b}}$ becomes positive
before the second critical configuration is reached
and remains positive with the further increase of $\sigma_0$.
Interestingly, for $\xi=4$ and greater,
at the second critical configuration $E_{\mathrm{b}}$ is negative.
The dependence of the binding energy (\ref{eq:ebind}) on $\sigma_0$
for some values of $\xi$ is shown in figure~\ref{fig:three};
one can see that for $\xi=4$,
there is a positive binding energy barrier
between the first and the second critical configuration,
while for sufficiently large values of $\xi$, this barrier disappears.
It should be emphasized, however,
that negative $E_{\mathrm{b}}$ at some equilibrium configuration
is a common characteristic of gravitationally bound states,
but does not by itself imply stability.


As outlined in the preceding section, in a given spacetime,
the photon spheres can be located
by examining the behaviour of the function $b(r_0)$ defined in (\ref{eq:br0}).
If $b(r_0)$ is monotonically increasing with $r_0 \ge 0$,
there is no photon sphere in the spacetime,
whereas if extrema exist, they correspond to photon spheres.
In this context, a more robust procedure
for locating the photon spheres can be formulated.
The condition for the extremum of $b(r_0)$ can be written as
   \begin{equation}
   0 = \frac{\mathrm{d}b}{\mathrm{d}r_0}
     = \mathrm{e}^{-\Phi} ( 1 - r \Phi' ) ,
   \end{equation}
where $\Phi'$ can be expressed in terms of the
mass function $m$ and the radial pressure $p$
using the well known equation
   \begin{equation}
   \Phi' = \frac{m + 4\pi r^3 p}{r^2 ( 1 - 2m/r )}
   \end{equation}
(in the non-relativistic regime, $\Phi$ is the gravitational potential).
It follows that the condition
for the existence of the photon sphere in the spacetime can be expressed as
   \begin{equation} \label{eq:pscond:mp}
   \frac{2m}{r} = \frac23 ( 1 - 4\pi r^2 p ).
   \end{equation}
The quantity on the lhs is known as the `compactness function'
and is always less than unity,
while the rhs is less than two-thirds if the radial pressure is positive.
It is interesting to observe that when one considers
a compact object of mass $M$ and surface radius $R$,
condition (\ref{eq:mu23}) for the photon sphere
in the vacuum segment of the spacetime
requires the surface compactness $2M/R$ to be greater than $2/3$;
here we find that, for positive radial pressure $p$,
the compactness function may satisfy the condition (\ref{eq:pscond:mp})
at values that are less than $2/3$.

\begin{figure}
\begin{center}
\includegraphics{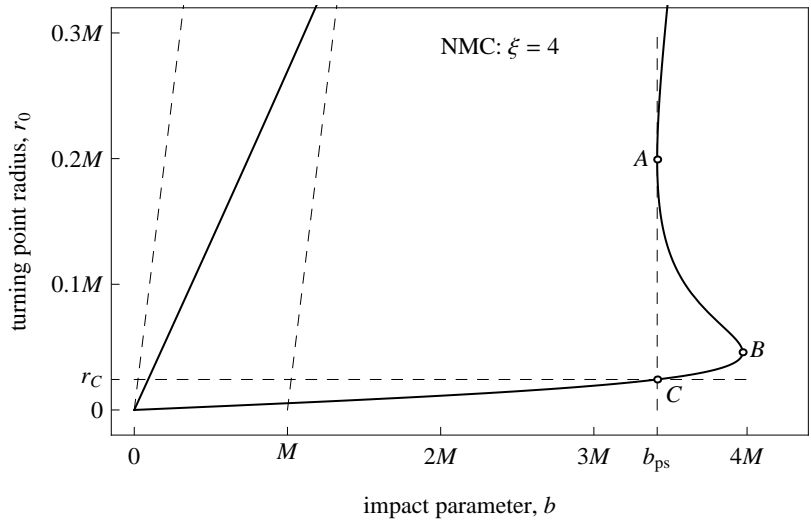}\\
\end{center}
\caption{\label{fig:four}
Radius of the turning point (or photon sphere) $r_0$
versus impact parameter $b$ for the null-geodesics in the spacetime
containing the boson star with the nonminimal coupling constant $\xi=4$
at the first (no photon spheres)
and second (photon spheres $A$ and $B$) critical configurations.
The impact parameter corresponding to the photon sphere $A$
is indicated by $b_{\mathrm{ps}}$,
and the area radius of the sphere containing the turning points
for null-geodesics with $b<b_{\mathrm{ps}}$ is indicated by $r_C$.
}
\end{figure}
\begin{figure}
\begin{center}
\includegraphics{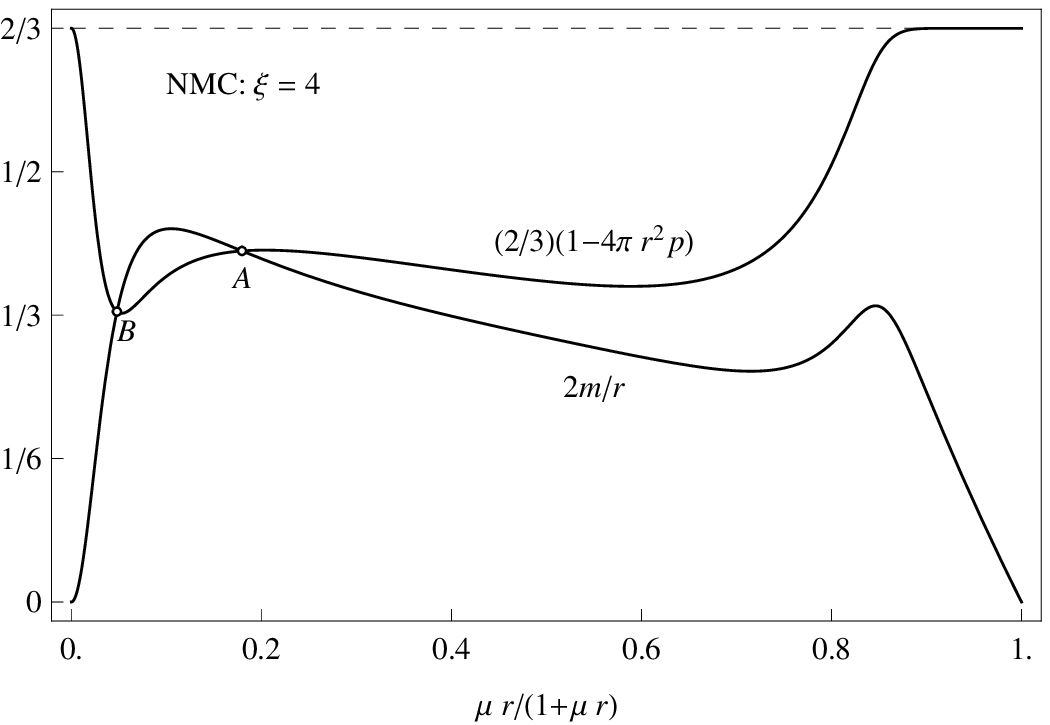}\\
\end{center}
\caption{\label{fig:five}
Plot locating the photon spheres $A$ and $B$ in the spacetime
containing the boson star with the nonminimal coupling constant $\xi=4$
at the second critical configuration as the solutions of (\ref{eq:pscond:mp}).
}
\end{figure}

In the range of values
of the non-minimal coupling constant $\xi$ that we considered
(see table \ref{tbl:one}),
we did not find photon spheres at or below
the value of $\sigma_0$ corresponding to the first critical configuration.
Interestingly, the photon spheres were found
in the second critical configurations for $\xi=0$ and above.
However, only at $\xi=4$ and above do we have negative binding energy
at the second critical configuration.
Figure \ref{fig:four} shows the relation among the turning point
radius $r_0$ and the impact parameter $b$ for the boson star with $\xi=4$.
Following the notation scheme introduced in the preceding section,
the outer (unstable) photon sphere is marked with $A$
and the inner (stable) photon sphere is marked with $B$.
The impact parameter corresponding to the photon sphere $A$
is denoted with $b_{\mathrm{ps}}$.
The area radius
of the sphere containing the turning points
of all null-geodesics with $b<b_{\mathrm{ps}}$ is denoted with $r_C$.
The layout of the points $A$, $B$ and $C$
for higher values of $\xi$ is very similar
to the layout shown in figure~\ref{fig:four}.
Figure~\ref{fig:five} shows how condition (\ref{eq:pscond:mp})
can be used to locate the photon spheres.


In the simple model of the spacetime with photon spheres,
constructed in the preceding section,
we found a strong degree of focusing of null-geodesics in the interior
of the compact object.
Here, as a measure of the shrinking of the bundle of null-geodesics
that are initially parallel at spatial infinity,
one can consider the ratio of the two proper areas defined as follows.
The first is the proper area
of the circular cross section of the part of the bundle at spatial infinity
containing all null-geodesics that (in the course of time)
pass through the photon sphere and enter its interior.
These geodesics must have the impact parameter $b<b_{\mathrm{ps}}$
so the proper area of the cross section is $A_\infty = b_{\mathrm{ps}}^2\pi$.
The second proper area is that of the sphere
containing the turning points of the geodesics
that enter the photon sphere.
As the area radius of this sphere is $r_C$,
its proper area is $A_C = 4r_C^2\pi$.
Defined in this way, the null-geodesics that pass through $A_\infty$
also pass through $A_C$, and vice versa.
The ratio of the two proper areas is
   \begin{equation} \label{eq:ratio}
   \frac{A_\infty}{A_C} = \frac{b_{\mathrm{ps}}^2}{4r_C^2}.
   \end{equation}
It can be interpreted as the ratio of the number of null-geodesics
passing through unit proper area at $r=r_C$ and at $r\to\infty$.
As it can be seen from the figures in table~\ref{tbl:one},
in the configurations of boson stars where we found photon spheres,
this ratio is by several orders of magnitude greater than unity.
If one also considered the number of null-particles (photons)
passing through unit proper area in unit proper time,
assuming steady flux at $r\to\infty$,
an additional factor of $\mathrm{e}^{-\Phi(r_C)} = b_{\mathrm{ps}}/r_C$
would enter due to the difference in clock ticking rates.


\begin{table}
\caption{\label{tbl:one}
Central values of the field variable $\sigma_0$,
the mass $M$ and the particle number $N$
of the first and second critical configurations of the boson star
obtained with a sequence of values of the nonminimal coupling constant $\xi$.
The impact parameters corresponding to the photon sphere $b_{\mathrm{ps}}$,
area radius of the photon sphere $r_A$
and the area radius $r_C$ of the sphere
containing the turning points for null-geodesics with $b<b_{\mathrm{ps}}$
are given for configurations where the photon sphere exists.
$M_{\mathrm{P}}$ denotes the Planck mass.
}
\begin{center}
\begin{tabular}{rllllll}
 $\xi$ & $(\sigma_0)_{\mathrm{crit.}}$
       & $M/(M_{\mathrm{P}}^2/\mu)$ & $N/(M_{\mathrm{P}}^2/\mu^2)$
       & $b_{\mathrm{ps}}/M$ & $r_A/M$ & $r_C/M$\\
\hline
  $16$ & $0.146287$ & $2.89786$  & $3.48375$  &           &            &             \\
       & $0.273464$ & $2.18529$  & $2.27896$  & $3.50852$ & $0.196796$ & $0.0233624$ \\
\hline
   $8$ & $0.203279$ & $2.05492$  & $2.44047$  &           &            &             \\
       & $0.383436$ & $1.54653$  & $1.59122$  & $3.47565$ & $0.197621$ & $0.0236546$ \\
\hline
   $4$ & $0.277568$ & $1.46454$  & $1.70347$  &           &            &             \\
       & $0.533207$ & $1.09811$  & $1.10413$  & $3.41338$ & $0.199297$ & $0.0242506$ \\
\hline
   $2$ & $0.365541$ & $1.05839$  & $1.19175$  &           &            &             \\
       & $0.729797$ & $0.78796$  & $0.764328$ & $3.29866$ & $0.202721$ & $0.0254954$ \\
\hline
   $0$ & $0.271059$ & $0.633001$ & $0.653003$ &           &            &             \\
       & $1.62527$  & $0.382621$ & $0.330516$ & $1.74432$ & $0.176402$ & $0.134695$  \\
\hline
\end{tabular}
\end{center}
\end{table}


\section{Conclusions \label{sec:concl}}


We have shown that photon spheres are present
in some configurations of the boson stars
constructed with the free massive scalar field
nonminimally coupled to gravity.
This implies the existence of null-geodesics 
that make arbitrarily many turns around the central region of the star
before escaping to spatial infinity.
Assuming the boson star is transparent to light,
this further leads to the possibility of formation of relativistic images
in the strong deflection regime of gravitational lensing.
With the finding of the photon spheres within the boson stars,
the scalar field appears to be a suitable matter model
for modeling of strong gravitational lenses
that do not involve singularities of black holes in the spacetime.
Another interesting feature of global (no black holes)
spherically symmetric spacetimes with photon spheres,
relative to those with black holes,
is the high degree of focusing
of the incident light into the small central region of the spacetime.
For the boson stars that we have considered,
we were able to give a measure of contraction of the incident bundle
of parallel null-geodesics in the form of the ratio (\ref{eq:ratio}).


However, one must not overlook the not-so-desirable
properties of the configurations of boson stars
in which photon spheres were found.
In the case of the minimally coupled field ($\xi=0$),
photon spheres were found in configurations
that are known to be dynamically unstable
and that have positive binding energy
(positive work is required to `compress' the bosons
with initially zero kinetic energy at spatial infinity
into the equilibrium configuration).
As we have shown, the unusual sign of the binding energy
of the configurations with photon spheres
can be reversed with the introduction
of sufficiently strong nonminimal coupling of the scalar field to gravity,
but the issue of instability remains open.
In this regard, it would be interesting
to carry out the linearized perturbation analysis
for the boson stars with a nonminimally coupled field
(e.g.\ like it has been done for $\xi=0$ in \cite{gleiserwatkins})
and if unstable modes are found in the configurations with photon spheres,
to obtain the timescales of the growing perturbations.
If these are found to be long relative to some time scale
reflecting the size of the spacetime region
in which gravitational lensing is taking place,
then the boson stars with photon spheres
would not be ruled out on the basis of instability
as models of massive dark objects
giving raise to phenomena of strong gravitational lensing.
The possible extension of this work could be
to explicitly study the angular pattern, magnifications and time delays
of the relativistic images formed by boson stars,
and compare them with the results that are obtained
when the massive dark objects in the centres of nearby galaxies
are modelled as black holes \cite{virbhadra2009,BozzaMancini2004}.


\vskip 1em


\noindent {\bf Acknowledgments:}
This work is supported by the Croatian Ministry of Science,
Education and Sports under the Project Nr 036-0982930-3144
and also by CompStar,
a Research Networking Programme of the European Science Foundation.



\providecommand{\href}[2]{#2}\begingroup\raggedright\endgroup


\end{document}